\documentstyle[12pt,epsfig,rotate]{article}

\newcommand{\bfr}{\begin{flushright}}
\newcommand{\beq}{\begin{equation}}
\newcommand{\efr}{\end{flushright}}
\newcommand{\eeq}{\end{equation}}
\begin{document}
\title{Damping of Collective Nuclear Motion and Thermodynamic
Properties of Nuclei beyond Mean Field\thanks{Part of the PhD thesis of Hong-Gang Luo;} \thanks{supported by DAAD, MFT, and GSI Darmstadt}}
\author{Hong-Gang  Luo$^{1,2}$,  W. Cassing$^1$ and  Shun-Jin  Wang$^{1,2}$
\\
1. Institut f\"ur Theoretische Physik, Universit\"at Giessen
\\
35392  Giessen, Germany
\\
2. Department of  Modern  Physics,  Lanzhou  University,
\\
Lanzhou  730000,  PR  China }
\maketitle
\begin{abstract}
The dynamical description of correlated nuclear motion is
based on a set of coupled
equations of motion for the one-body density matrix $\rho (11';t)$ and the
two-body correlation function $c_2(12,1'2';t)$, which is obtained from the
density-matrix hierarchy beyond conventional mean-field  approaches
by truncating 3-body correlations. The resulting equations
nonperturbatively describe
particle-particle collisions (short-range correlations) as well as particle-hole interactions
(long-range correlations).
Within a basis of time-dependent Hartree-Fock states these equations
of motion are solved for collective vibrations of $^{40}Ca$ at several finite
thermal excitation energies corresponding to temperatures  $T=0-6$ MeV.
Transport coefficients for friction and diffusion are extracted from the
explicit solutions in comparison to the  solutions of the associated
 TDHF, VUU, Vlasov or damped quantum oscillator equations of motion. We find that
the actual magnitude of the transport coefficients is strongly influenced
by partlicle-hole correlations at low temperature which generate large
fluctuations in the nuclear shape degrees of freedom. Thermodynamically,
 the specific heat and the entropy of the system as a function of temperature
does not differ much from the mean-field limit except for a bump in the
specific heat around
 $T\simeq 4$ MeV  which we attribute to the melting of shell effects in the
correlated system.
\end{abstract}

\vspace{1cm}

\vspace{0.5cm}

\newpage
\section{Introduction}

The damping of nuclear collective motion and the role of short-range and
long-range nuclear correlations \cite{B1,B2} in the relaxation still is left
as an open problem from the fully microscopic point of view. Whereas at
intermediate  energy heavy-ion collisions the system is essentially damped by
in-medium nucleon-nucleon collisions within the time-dependent mean field
\cite{r1,r2,Bona}, the relaxation of giant resonances at low excitation energy
appears to be dominated by residual 2 particle -- 2
hole (2p-2h) correlations \cite{r3,r4} that are not incorporated in standard
transport theories \cite{r1,r2,Bona}.
Though the latter approaches have proven to quite
reliably describe one-body observables in heavy-ion reactions, they fail
in describing fluctuation phenomena like 'multifragmentation'. Thus Langevin
forces -  that are related to fluctuations in the particle collision number
or in momentum space -
 have been proposed in addition to the Vlasov-Uehling-Uhlenbeck
(VUU) dynamics  \cite{r20} - \cite{r28}.  Since the various approaches
involve quite different assumptions about the dynamical origin and the size
of 'fluctuations', a more rigorous analysis within the framework of a
nonperturbative quantum two-body theory appears necessary.

  More specifically, nuclear giant resonances of hot nuclei have been
studied in detail in both experimental \cite{exp1,exp2,20b,20c}
and theoretical nuclear physics \cite{20d} - \cite{28b}
to learn about the nuclear collectivity and the lifetime of these modes.
 The stability of the damping
widths against temperature, the effects of shape degrees of freedom on
the damping process, the role
played by particle-hole (long-range) correlations, and the distinction among the damping
processes in
different dynamical models have presented open and interesting  problems.
Though  several authors have
addressed these problems within various approaches
\cite{r3,r4,20d,V1,V2,V3,T1,T2,T3,28a,28b,r10,s2,r30,31b},
many microscopic aspects
remain unclear. It is the purpose of this paper to investigate
these problems from a systematical point of view and in a more detailed way
especially in a comparison of quantum mechanical and semiclassical approaches.

In Section 2 we briefly review the underlying  microscopic approach
obtained from the correlation dynamical truncation scheme \cite{r5}
and present the
approximations performed for the actual calculations. Section 3 is devoted
to a presentation of the microscopic results in case of isoscalar
quadrupole motion of $^{40}Ca$ and to the extraction of transport
coefficients from the mean values and dispersions of the quadrupole operator
$Q$ and its time derivative $\dot{Q}$. The results will also be compared to the
mean-field limit TDHF as well as to the semiclassical limits of VUU and
solutions based on the Vlasov equation.
Furthermore, in Section 4 the specific heat,  entropy and free energy of the
nucleus $^{40}Ca$ will be evaluated as a function of temperature in the
correlated two-body theory in comparison to the mean-field limit.
 A summary and discussion in  Section 5 concludes this study.

\section{Nuclear quantum correlation dynamics (NQCD)}

The dynamical description of the nuclear many-body problem in the
non-relativistic limit is based on coupled equations of motion for the one-body
density matrix $\rho (11';t)$ and the two-body correlation function
$c_2(12,1'2';t)$ in the equal time limit \cite{r5} - \cite{r9};
$$
{\rm i}\partial /\partial t\ \rho (11';t) =
 [t(1) + U(1;t)] \rho(11';t) - \rho(11';t) [t(1') + U^{\dagger}(1';t)]
$$
\beq
\label{2.1}
+ Tr_{(2=2')} [v(12) c_2(12,1'2';t) - c_2(12,1'2';t) v(1'2')]
\eeq
where the two-body density matrix is written as
$ \rho _2(12,1'2';t) =
\rho _{20}(12,1'2';t) + c_2(12,1'2';t) $  with $
\rho _{20}(12,1'2';t) =
{\cal A}_{12}\rho (11';t)\rho (22';t)$ .
In (\ref{2.1}) $t(i)$ denotes the kinetic
energy of particle $i$ while $v(ij)$ is the bare two-body interaction. The time
evolution of $c_2$ is determined by
$$
{\rm i}\partial /\partial t\ c_2(12,1'2';t) =
[t(1)+t(2)+U(1;t)+U(2;t)]c_2(12,1'2';t)
$$
$$
- c_2(12,1'2';t)[t(1')+t(2')+U^{\dagger}(1';t)+ U^{\dagger}(2';t)]
$$
\beq
\label{2.2a}
+ [Q^=_{12}v(12)\rho _{20}(12,1'2';t)-\rho _{20}(12,1'2';t)v(1'2')
Q^{=\dagger}_{1'2'}]
\eeq
\beq
\label{2.2b}
+ [Q^=_{12}v(12)c_2(12,1'2';t) - c_2(12,1'2';t)v(1'2')Q^{=\dagger}_{1'2'}]
\eeq
$$
 + Tr_{(3=3')}
 [v(13){\cal A}_{13}{\cal A}_{1'2'} - v(1'3'){\cal A}_{1'3'}{\cal A}_{12}]
\rho (11';t)c_2(23,2'3';t)
$$
\beq
\label{2.2c}
 + Tr_{(3=3')}
 [v(23){\cal A}_{23}{\cal A}_{1'2'} - v(2'3'){\cal A}_{2'3'}{\cal A}_{12}]
\rho (22';t)c_2(13,1'3';t)
\eeq
\beq
\label{2.2d}
+ Tr_{(3=3')}
 \{[v(13)+v(23)]c_3(123,1'2'3';t)-c_3(123,1'2'3';t)[v(1'3')+v(2'3')]\}.
\eeq
\noindent
In equations (\ref{2.1}) - (\ref{2.2d}) $Tr_{(i=i')}$
 stands for a summation over spin $\sigma (i)$ and isospin $\tau (i)$ as
well as an integration over $d^3r_i$ with $i=i'$, while ${\cal A}_{ij}$ is the
antisymmetrization operator for fermions defined by
${\cal A}_{ij} = 1 - P_{ij}$
and $P_{ij}$ is the permutation operator between particle $i$ and $j$. In
(\ref{2.1}) -  (\ref{2.2d}) we have, furthermore,
introduced the mean-field potential

\beq
\label{2.3}
U(i;t) =  Tr_{(3=3')}  \{v(i3){\cal A}_{i3}\rho (33';t)\}
\eeq
and the Pauli-blocking operator for $p-p$ and $h-h$ interactions

\beq
\label{2.4}
Q^=_{ij} = 1 - Tr_{(3=3')} (P_{i3}+P_{j3})\rho (33';t)
\eeq
In order to close (\ref{2.1}) - (\ref{2.2d})
the three-body correlation term (\ref{2.2d}) will
be dropped in the following. For an explicit discussion of (\ref{2.2d}) in
the context of trace conservation laws we refer the reader to Ref. \cite{r9}.

\subsection{Time-Dependent Density-Matrix Theory (TDDM)}

Because of technical limitations the set of coupled
equations for the one-body density matrix $\rho(11';t)$ (\ref{2.1}) and
the two-body correlation function $c_2(12,1'2';t)$ (\ref{2.2a}) - (\ref{2.2c})
 can, at present, only
be solved accurately for moderate excitations of the nuclear system.
In this case one can expand both $\rho$
and $c_2$ in terms of single-particle states $\psi_{\alpha}$
fulfilling the TDHF equations,
\beq
\label{2.5}
({\rm i}\ \frac{\partial}{\partial t} - h(1)) \psi_{\alpha} (1,t) = 0
\eeq
according to
$$
\rho(11';t)=\sum_{{\alpha},{\beta}}n_{{\alpha}{\beta}}(t) \
\psi_{\beta}^{*}(1',t)\psi_{\alpha}(1,t)
$$
\beq
\label{2.6}
c_2(12,1'2';t)=\sum_{{\alpha},{\beta},{\alpha'},{\beta'}}C_{\alpha
\beta \alpha ' \beta '}(t) \
\psi_{\alpha}(1,t)\psi_{\beta}(2,t)\psi^*_{\alpha
'}(1',t)\psi^*_{\beta '}(2',t),
\eeq
where the operator $h(i)=t(i)+U(i)$ in (\ref{2.5})
is the one-body hamiltonian, i.e. the sum
of a kinetic energy term and the selfconsistent mean field (\ref{2.3}) .

The equations of motion for the occupation
matrix $n_{\alpha \beta}(t)$ and $C_{\alpha \beta \alpha' \beta'}(t)$ then read
\cite{r4,r10,r30}
\beq
\label{2.7}
{\rm i}\ \frac{\partial}{\partial t}{n_{\alpha \beta}}=
\sum_{\gamma \delta \sigma}
\{ \langle \alpha \sigma|v|\gamma \delta\rangle  C_{\gamma \delta \beta \sigma}
-C_{\alpha \delta \gamma \sigma}\langle \gamma \sigma|v|\beta \delta\rangle \}
\eeq
and
\beq
\label{2.8}
{\rm i}\ \frac{\partial}{\partial t}{C_{\alpha \beta \alpha ' \beta '}}=
B_{\alpha \beta \alpha ' \beta '}+
P_{\alpha \beta \alpha ' \beta '}+
H_{\alpha \beta \alpha ' \beta '},
\eeq
where
$$
B_{\alpha \beta \alpha ' \beta '}=\sum_{\lambda_1 \lambda_2
\lambda_3 \lambda_4}
\langle \lambda_1 \lambda_2|v|\lambda_3 \lambda_4\rangle _{\cal A}
$$
\beq
\label{2.9}
\cdot \{(\delta_{\alpha \lambda_1}-n_{\alpha \lambda_1})
(\delta_{\beta \lambda_2}-n_{\beta \lambda_2})
n_{\lambda_3 \alpha '}n_{\lambda_4 \beta '}
-n_{\alpha \lambda_1}n_{\beta \lambda_2}
(\delta_{\lambda_3 \alpha '}-n_{\lambda_3 \alpha '})
(\delta_{\lambda_4 \beta '}-n_{\lambda_4 \beta '})\}
\eeq
represents the lowest order contribution of collisions in the
particle-particle channel (Born approximation), while
the term $P$ represents the higher order particle-particle (and hole-
hole) contributions,
$$
P_{\alpha \beta \alpha ' \beta '}=
\sum_{{\lambda _1}{\lambda _2}{\lambda _3}{\lambda _4}}
\langle {\lambda _1}{\lambda _2}|v|{\lambda _3}{\lambda _4}\rangle
[\delta_{\alpha \lambda_1} \delta_{\beta \lambda_2} C_{\lambda_3 \lambda_4
\alpha' \beta'}
-\delta_{\lambda_3 \alpha'} \delta_{\lambda_4 \beta '} C_{
\alpha \beta \lambda_1 \lambda_2 }
$$
\beq
\label{2.10}
-\delta_{\alpha \lambda_1} n_{\beta \lambda_2} C_{\lambda_3 \lambda_4
\alpha ' \beta '}
-\delta_{\lambda_2 \beta} n_{\alpha \lambda_1} C_{
\lambda_4 \lambda_3 \beta ' \alpha ' }
+\delta_{\lambda_3 \alpha ' } n_{\lambda_4 \beta '} C_{
\alpha \beta \lambda_1 \lambda_2 }
+\delta_{\lambda_4 \beta '} n_{\lambda_3 \alpha '} C_{
\alpha \beta \lambda_1 \lambda_2  }].
\eeq
The last term
$$
H_{\alpha \beta \alpha ' \beta '}=
\sum_{{\lambda _1}{\lambda _2}{\lambda _3}{\lambda _4}}
\langle {\lambda _1}{\lambda _2}|v|{\lambda _3}{\lambda _4}\rangle
$$
$$
[\delta _{\alpha \lambda _1}(n_{{\lambda _3}{\alpha '}} C_{{\beta}{\lambda _4}
{\beta '}{\lambda_2}}-n_{{\lambda _3}{\beta '}}C_{{\beta}{\lambda _4}{\alpha '}
{\lambda_2}}
-n_{\lambda _4 \alpha '}C_{\lambda _3 \beta \lambda _2 \beta '}
-n_{\lambda _4 \beta '}C_{\lambda _3 \beta \alpha ' \lambda _2})
$$
$$
+{\delta}_{{\beta}{\lambda_2}}(n_{{\lambda _4}{\beta '}}
C_{{\alpha }{\lambda _3}{\alpha '}{\lambda_1}}
-n_{{\lambda _4 }{\alpha '}}C_{{\alpha }{\lambda _3 }{\beta ' }{\lambda _1}}
-n_{{\lambda _3 }{\beta '}}C_{{\alpha }{\lambda _4 }{\alpha ' }{\lambda _1}}
-n_{{\lambda _3 }{\alpha '}}C_{{\alpha }{\lambda _4 }{\lambda _1}{\beta '}})
$$
$$
-{\delta}_{{\beta ' }{\lambda _4}}(n_{{\beta}{\lambda _2}}
C_{{\alpha }{\lambda _3}{\alpha ' }{\lambda_1}}
-n_{{\alpha }{\lambda _2}}C_{{\beta }{\lambda _3 }{\alpha '}{\lambda _1}}
-n_{{\beta }{\lambda _1}}C_{{\alpha }{\lambda _3 }{\alpha ' }{\lambda _2 }}
-n_{{\alpha }{\lambda _1}}C_{{\beta }{\lambda _3 }{\lambda _2 }{\alpha}})
$$
\beq
\label{2.11}
-{\delta}_{{\alpha ' }{\lambda _3}}(n_{{\alpha}{\lambda _1}}
C_{{\beta }{\lambda _4}{\beta ' }{\lambda_2}}
-n_{{\alpha }{\lambda _2}}C_{{\alpha }{\lambda _4 }{\beta '}{\lambda _2}}
-n_{{\alpha }{\lambda _2}}C_{{\beta }{\lambda _4 }{\beta ' }{\lambda _1 }}
-n_{{\beta }{\lambda _2}}C_{{\alpha }{\lambda _4 }{\lambda _1 }{\beta '}})]
\eeq
is the contribution to the equations of motion in the particle-hole channel.
The set of coupled equations (\ref{2.7}) and (\ref{2.8})
provides a non-perturbative
description of large amplitude nuclear motion denoted as Time-Dependent
Density-Matrix ({\bf TDDM}) theory which takes into account collisions
among nucleons to all orders in the interaction and thus superseeds the
quantum approaches in Refs. \cite{28b,35b} that are limited to the
Born approximation, i.e. retaining only $B_{alpha \beta \alpha' \beta'}$
in Eq. (\ref{2.8}). We note that the
theory fulfills the conservation of particle number, momentum, and energy
\cite{r10}.

\subsection{Dispersion of one-body operators}

Since TDDM is a consistent two-body theory it allows to evaluate one-body
as well as two-body observables. Of specific interest are the mean
values and the fluctuation properties of collective one-body operators,
for which we give the microscopic expressions. Note that contrary to Refs.
\cite{20d,31b,35c} no explicit equations of motion for a collective degree of
freedom have to be set up since Eqs. (10) and (11) already describe the
full dynamics.

The quadrupole operator (as well as any one-body operator $O$) is defined as
\beq
\label{2.12}
O =\sum_{\alpha,\beta} \langle \alpha|O|\beta\rangle a^{+}_{\alpha}a_{\beta},
\eeq
where $a^{+}_{\alpha}$ and $a_{\beta}$ are creation and annihilation
operators of a nucleon.
Our definition of the quadrupole operator $Q_2({\bf r}) \equiv Q({\bf r})$
is explicitly
\beq
\label{2.13}
Q({\bf r})={1\over 2} \sqrt{{5\over {4\pi}}}(2z^{2}-x^{2}-y^{2})
\eeq
which gives for the time derivative
\beq
\label{2.14}
\dot{Q}({\bf r},{\bf p}) = {-{\rm i} \over {\hbar}} [Q,h] = \sqrt{{5\over{4\pi}}} {1 \over {2m}}
(2 z p_z + 2 p_z z - x p_x - p_x x - y p_y - p_y y).
\eeq
The inertia mass operator of the quadrupole motion, $M({\bf r})$, is
\beq
\label{2.15}
\frac{1}{M({\bf r})}= -\frac{1}{\hbar^{2}}[ Q({\bf r}),[Q({\bf r}), h ]]=
-\frac{{\rm i}}{\hbar}[Q({\bf r}),\dot{Q}({\bf r},{\bf p})]=\frac{5}{4\pi m}
\left(4z^{2} +x^{2} +y^{2}\right).
\eeq
Eq. (\ref{2.15}) presents not only an expression for the quadrupole mass operator
but  also a more general commutation relation  for  the quadrupole
operator $Q({\bf r})$ and its velocity operator $\dot{Q}({\bf r},{\bf p})$,
in which the constant $i\hbar$ is replaced by the operator
${i \hbar}/{M({\bf r})}$. This is equivalent to a collective momentum operator
$P = M({\bf r}) \dot{Q}({\bf r, p})$.

From the commutation relation (\ref{2.15}) one can derive the
corresopnding uncertainty relation as,
\beq
\label{2.15a}
\sigma^{2}_{Q} \sigma^{2}_{\dot{Q}} \geq \left(\frac{\hbar}{2}\langle  \frac{1}
{M({\bf r})}\rangle  \right)^{2},
\eeq
or
\beq
\label{2.15b}
\sigma^{2}_{Q} M^{2}_{Q} \sigma^{2}_{\dot{Q}} \geq \left(\frac{\hbar}{2}\right)^{2},
\eeq
with the definition of the quadrupole mass $1/M_{Q}=
\langle  1/M({\bf r})\rangle  $,  and with $\sigma^{2}_{Q}$ and $\sigma^{2}_{\dot{Q}}$
defined by Eq. (\ref{2.16}) below. Eq. (\ref{2.15b}) will be checked as a
boundary condition in our following numerical calculations explicitly.

The mean value and the dispersion of a one-body operator $O$ are given by:
\beq
\label{mean}
\langle O\rangle =\sum_{\alpha,\beta} \langle  \alpha|O|\beta\rangle \langle  a^{+}_{\alpha}a_{\beta}\rangle
=  \sum_{\alpha,\beta} \langle \alpha|O|\beta\rangle  n_{\beta \alpha}
\eeq
$$
\sigma^{2}_{O}
=\sum_{\alpha,\alpha ',\beta,\beta'} \langle \alpha|O|\beta\rangle \langle \alpha'|O|\beta'\rangle
\langle a^{+}_{\alpha}a_{\beta}a^{+}_{\alpha'}a_{\beta'}\rangle -\langle O\rangle ^{2}
$$
$$
=\sum_{\alpha,\alpha ',\beta,\beta'} \langle \alpha|O|\beta\rangle \langle \alpha'|O|\beta'\rangle
(\delta_{\alpha' \beta}\langle a^{+}_{\alpha}a_{\beta'}\rangle
+\langle a^{+}_{\alpha}a^{+}_{\alpha'}a_{\beta'}a_{\beta}\rangle )-\langle O\rangle ^{2}
$$
$$
=\sum_{\alpha, \beta, \beta '} \langle \alpha|O|\beta\rangle \langle \beta|O|\beta '\rangle
n_{\beta '\alpha}
+  \sum_{\alpha,\alpha ',\beta,\beta'} \langle \alpha|O|\beta\rangle \langle \alpha'|O|\beta'\rangle
(C_{\beta \beta' \alpha \alpha'} - n_{\beta \alpha'}n_{\beta' \alpha})
$$
\beq
\label{2.16}
=\sum_{\alpha, \beta '} \langle \alpha|O^{2}|\beta '\rangle  n_{\beta'\alpha}
+ \sum_{\alpha,\alpha',\beta,\beta'}\langle \alpha|O|\beta\rangle \langle \alpha'|O|\beta'\rangle
(C_{\beta \beta' \alpha \alpha'}- n_{\beta \alpha'} n_{\beta' \alpha}),
\eeq
where the matrix elements $n_{\alpha \beta}$ and
$C_{\alpha \beta \alpha' \beta'}$ depend on time explicitly.
In case of TDHF we simply have $C_{\beta \beta' \alpha \alpha'}=0$ and $n_{\alpha \beta}=\delta
_{\alpha \beta}n_{\alpha}$ where $n_{\alpha}$ denotes the occupation probability
of the single-particle state $\psi_{\alpha}$. This gives
\beq
\label{mean2}
\langle O\rangle =\sum_{\alpha} \langle \alpha|O|\alpha\rangle n_{\alpha} ,
\eeq
$$
\sigma^{2}_{O} = \sum_{\alpha,\beta}
\langle \alpha|O|\beta\rangle (1 - n_{\beta})\langle \beta|O|\alpha\rangle   n_{\alpha}
$$
\beq
\label{2.17}
=\sum_{\alpha} \langle \alpha|O^{2}|\alpha\rangle  n_{\alpha}-\sum_{\alpha,\beta}
\langle \alpha|O|\beta\rangle \langle \beta|O|\alpha\rangle  n_{\beta} n_{\alpha}.
\eeq
We note that the summation over $\beta$ in the first terms of
Eqs. (\ref{2.16}) and (\ref{2.17}) should be taken
in the whole Hilbert space ($\sum_\beta |\beta><\beta| =1 $)  so that it
 yields an identity in the whole Hilbert
space. This observation is important since the
numerical calculations are carried out in a truncated Hilbert space.

\subsection{Technical approximations}

In our actual calculations we approximate the interaction $v$ appearing
in the mean-field potential (\ref{2.3})
by a Bonche-Koonin-Negele (BKN) \cite{r15}
 force and use $v(12)$=V$_0\delta({\bf r}_1- {\bf r}_2)$
 for the residual interaction appearing in (\ref{2.9})-(\ref{2.11})
with V$_0 = -300$
MeV fm$^3$ for technical reasons (see below) as in Refs. \cite{r4,r30,Alfred}.
We have solved the coupled equations
 (\ref{2.7}) and (\ref{2.8}) for the case of
the isoscalar monopole or quadrupole vibration of
 $^{40}Ca$ as a function of temperature since the dipole mode was found in Ref. [7]
to be dominated by mean-field dynamics.
To carry out calculations at various temperatures we describe
the initial occupation numbers by appropriate Fermi distributions
(characterized by a temperature $T$) and propagate the system in time.
The set of single-particle levels used in the calculations
include the 1s, 1p, 2s, 1d, 2p, and 1f shell.   The correlated
state, which is generated by an adiabatic switching on of the residual
interaction \cite{Alfred,Peter},
 is then boosted at a certain time (typically after $2 \cdot 10^{-22} s$)
by applying appropriate phase factors to the wavefunctions $\psi_\beta$
proportional to a strength
factor $\alpha$. In this way the system acquires a well defined
collective energy proportional to $ \alpha^2$.
The set of equations (\ref{2.5}) - (\ref{2.8}) is then integrated
 in time with timestep-size
$\Delta t = 0.5 \cdot 10^{-23} s $. We note that
the total energy and number of particles are conserved well throughout
the timescales of interest here.

\subsection{ Specific heat, entropy, and free energy }

  The energy of the system reads for the present Skyrme interaction as well as
the residual $\delta-$interaction:
$$
E= \int \epsilon ({\bf r})\  d^{3}{\bf r} = \int d^{3}r \
[ \frac{\hbar^{2}}{2m}(\tau_n+ \tau_p)+
\frac{t_0}{2}(2\rho_n\rho_p+\frac{1}{2} (\rho_n^{2}+\rho_p^{2}))
$$
$$
+\frac{t_3}{4}(\rho_n^{2}\rho_p+\rho_n\rho_p^{2}) ]
+\frac{\upsilon_L}{2}[ E_y(\rho_n,\rho_n)
+E_y(\rho_p,\rho_p) ]+\upsilon_U E_y(\rho_n,\rho_p)
$$
\beq
\label{2.18}
+\frac{e^2}{2} \int \int d^{3}{\bf r} d^{3}{\bf r'} \
\frac{\rho_p({\bf r})\rho_p({\bf r'})}{\mid {\bf r }-{\bf r'}\mid}
+ \frac{V_0}{2} \int   C_2({\bf r},{\bf r};{\bf r},{\bf r}) \ d^{3}{\bf r},
\eeq
where
\beq
\label{2.19}
E_y(\rho_q, \rho_{q'})=\int \int d^{3}{\bf r} d^{3}{\bf r'} \
\frac{\exp(-\mu\mid {\bf r}-{\bf r'}\mid )}{\mu \mid
{\bf r}-{\bf r'}\mid} \rho_q({\bf r})\rho_{q'}({\bf r'})
\eeq
results from the finite range Yukawa force \cite{r15}.
The total energy of the system  - after adiabatically switching on the
residual interactions - depends on the initial thermal excitation
energy characterized by a temperature $T$, from which one can calculate
numerically the specific heat as
\beq
\label{2.20a}
C_v(T)= \frac{\partial E(T)}{\partial T}.
\eeq
The entropy  for the many-body system, which should not be mixed up with
the one-body entropy, then is given by
\beq
\label{2.20b}
S(T)=\int_0^T \frac{C_v(T')}{T'} dT'
\eeq
and the free energy follows as
\beq
\label{2.20c}
F(T)=E-TS.
\eeq
Furthermore, the  fluctuation  in energy of the system can be
evaluated as,
\beq
\label{2.21}
\sigma^2_H= \langle (H-\langle H\rangle )^2\rangle =\langle H^2\rangle -E^2=T^2C_v,
\eeq
which can be considered as an order parameter for a phase transition.

\subsection{The semiclassical limit}

As shown in detail in Ref. \cite{r2} the coupled equations of motion (\ref{2.1})
- (\ref{2.2b}), i.e. excluding (4),
transform to a Vlasov-Uehling-Uhlenbeck (VUU) equation in
the semiclassical limit when adopting an on-shell (Markov) approximation
for the $NN$-interactions and expanding up to the first order in the derivative
operator $\hbar^{-1} \nabla_r \cdot \nabla_p$ with respect to the particle
phase-space variables ${\bf r}$ and ${\bf p}$. This gives the familiar VUU
equation (for momentum-independent mean-fields)
\beq
\label{vuu}
\{\frac{\partial}{\partial t} + \frac{{\bf p}}{E} \cdot \nabla_r
- \nabla_r U({\bf r}) \cdot \nabla_p\} f({\bf r}, {\bf p};t)
= I_{coll}({\bf r}, {\bf p};t),
\eeq
where the collision term is given by \cite{r2}
$$
I_{coll}({\bf r}, {\bf p};t) = \frac{4}{(2\pi)^3} \int d^3 p_2 \int d\Omega
\ |v_{12}| \frac{d \sigma}{d \Omega}({\bf p},{\bf p}_2)
$$
$$
\times (f({\bf r}, {\bf p}_3;t) f({\bf r}, {\bf p}_4;t)
(1 - f({\bf r}, {\bf p};t))(1- f({\bf r}, {\bf p}_2;t))
$$
\beq
\label{coll}
- f({\bf r}, {\bf p};t) f({\bf r}, {\bf p}_2;t)
(1 - f({\bf r}, {\bf p}_3;t))(1- f({\bf r}, {\bf p}_4;t)).
\eeq
In Eq. (\ref{coll}) ${\rm d} \sigma/{\rm d} \Omega $ denotes the differential cross section
for the on-shell scattering process $1 + 2 \rightarrow 3 + 4$ while $|v_{12}|$
is the relative velocity of the colliding nucleons; the outgoing momenta
${\bf p}_3$ and ${\bf p}_4$ are fixed by energy- and momentum conservation
except for the scattering angle $\Omega$ in their center-of-mass.
When neglecting the collision term (\ref{coll}), the Vlasov equation (\ref{vuu})
is equivalent to the statement, that the total time derivative of the one-body
phase-space density $f({\bf r}, {\bf p}; t)$ vanishes, i.e ${\rm d}/{\rm d}t f$ = 0. The
Vlasov equation thus corresponds to the semiclassical limit of TDHF.
The VUU equation (\ref{vuu}) is conventionally \cite{r1,r2} solved within the
testparticle method and has been exploited extensively for the nonequilibrium
description of proton-nucleus or nucleus-nucleus collisions at intermediate
\cite{r2} and high energies \cite{WC}. Here we will compare the results of
the semiclassical VUU or Vlasov equation with those for the quantum
mechanical theories TDDM and TDHF, respectively.

\subsection{The damped harmonic oscillator in quantum physics}

For later comparison and interpretation of the numerical results we recall the
equations of motion for a damped harmonic quantum oscillator (QHO) with
respect to fluctuations in the collective coordinates $Q$ and $\dot{Q}$
(cf. Refs. \cite{20d,31b,35c}):
\beq
\label{Q1}
\frac{{\rm d}}{{\rm d}t} \sigma^2_Q = 2 \sigma_{Q \dot{Q}}
\eeq
\beq
\label{Q2}
\frac{{\rm d}}{{\rm d}t} \sigma_{Q \dot{Q}} = - 2 \gamma_Q \sigma_{Q \dot{Q}}
+ \frac{\Omega^2}{M_Q}
\sigma^2_Q
\eeq
\beq
\label{Q3}
\frac{{\rm d}}{{\rm d}t} \sigma^2_{\dot{Q}} = - 2 \gamma_Q \sigma^2_{\dot{Q}} +
\frac{\Omega^2}{M_Q} \sigma_{Q \dot{Q}} + D_{QQ}(T),
\eeq
where the fluctuations are defined as
$$\sigma^2_Q = \langle Q^2\rangle - \langle Q\rangle^2;$$
$$\sigma_{Q \dot{Q}} = \langle Q \dot{Q}\rangle - \langle Q\rangle \langle \dot{Q}\rangle;
$$
\beq
\label{diffus}
\sigma^2_{\dot{Q}} = \langle \dot{Q}^2\rangle - \langle \dot{Q}\rangle^2.
\eeq
In Eq. (\ref{Q2}) $\Omega$ denotes the collective frequency, $\gamma_Q$ a collective
friction coefficient, $M_Q$ the collective mass parameter, whereas $D_{QQ}(T)$ is
the diffusion coefficient given by
\beq
\label{diffu}
D_{QQ}(T) = \frac{\hbar\Omega \gamma_Q}{M_Q}\coth\frac{\hbar\Omega}{2T}.
\eeq
Note that only for large temperatures $T \gg \hbar \Omega$ the Einstein relation
\beq
\label{albert}
D_{QQ} = 2 \gamma_Q T/M_Q
\eeq
is obtained. In view of (small amplitude)
nuclear collective quadrupole motion we have $T < \hbar \Omega$
such that relation (\ref{albert}) does not apply. When using $P \equiv
M_Q {\dot Q}$ instead of ${\dot Q}$ the diffusion coefficient (\ref{diffu})
or (\ref{albert}) picks up an additional factor $M^2_Q$.

\section{Numerical results}

As a first test we study the 'hole-strength' distribution of $^{40}Ca$
within the nonperturbative limit TDDM by integrating the coupled
equations (\ref{2.5}), (\ref{2.7}), and (\ref{2.8}) in time with the
initial condition $C_{\alpha \beta \alpha' \beta'} = 0$, which
corresponds to the respective Hartree-Fock groundstate. The correlated
state is then generated in line with the Gell-Mann-Low theorem by
adiabatically switching on the residual interaction as
\beq
\label{switch}
V(t)= V_0\,[1-\exp(-t/\tau)] ,
\eeq
where $\tau \simeq 5 \cdot 10^{-22} sec$ was
found to be sufficiently large to achieve convergence (cf. also Ref. \cite{Alfred}).
The numerical
solutions for the occupation numbers $n_{\alpha}(t)$  deviate from 0
 and 1 and slightly oscillate in time. By averaging over time (for t
$\geq 10^{-21} sec$) we obtain average occupation numbers $\langle
n_{\alpha}\rangle $ which are shown in Fig.\ 1 as a function of the
time-averaged single-particle energies $\epsilon_{\alpha}$. The solid line
presents a smooth fit through the discrete points for orientation. The
latter one is remarkably close to the experimental data from Ref.
\cite{lap93} which indicates that the proper strength of the residual
correlations is well described by TDDM for $V_0 = - 300$ MeV fm$^3$,
which we consider as fixed from now on.

\begin{figure}[b]
\centerline{\psfig{figure=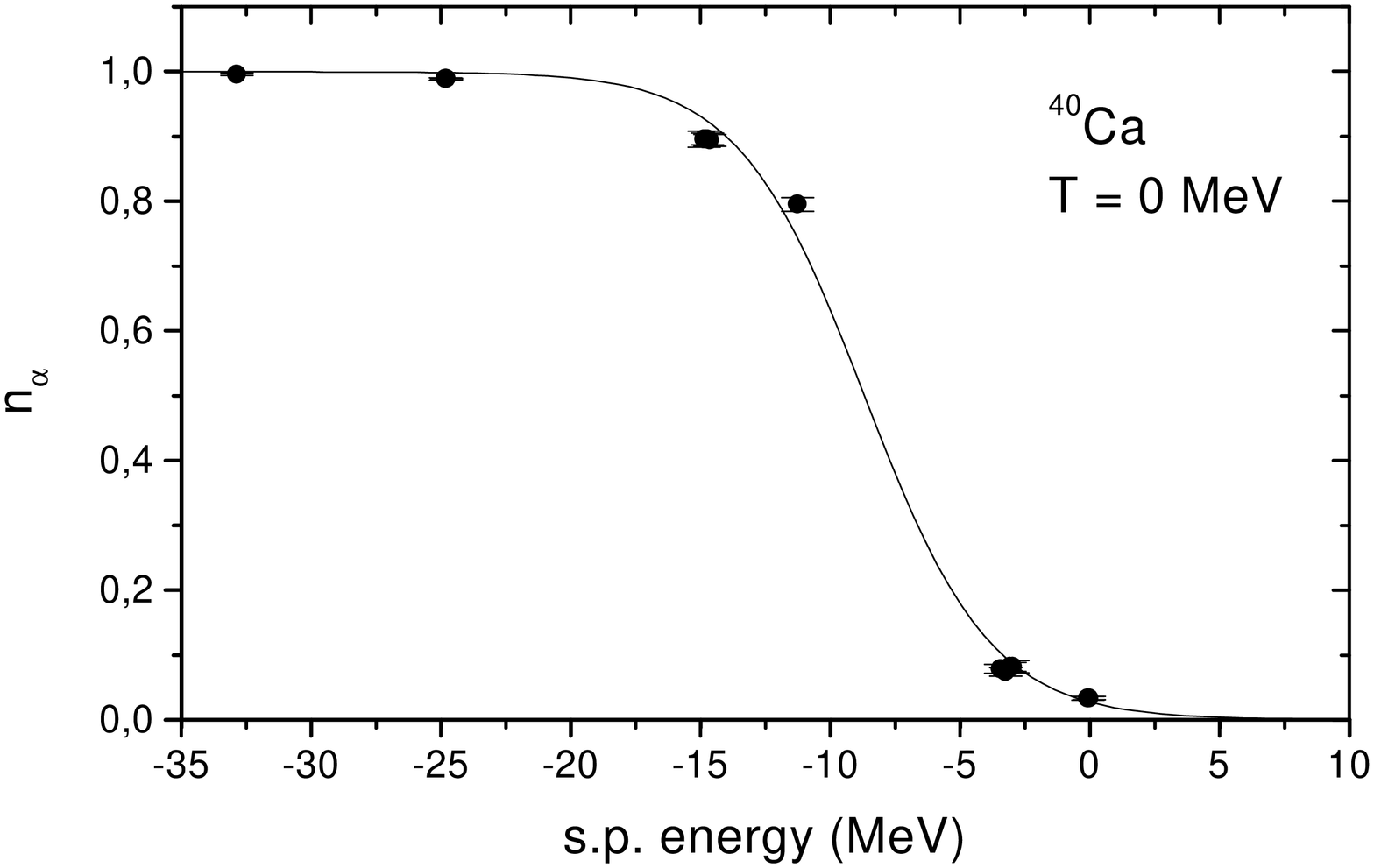,width=10cm,angle=-90}}
\vspace{1.5cm}
{\small Fig.~1: The time averaged occupation numbers $n_{\alpha}$
for $^{40}Ca$ as a
function of the s.p. energies $\epsilon_{\alpha}$ at T = 0 MeV in
the limit TDDM.}
\label{fig1}
\end{figure}


We note in passing that the average occupation numbers have no direct
counter part in the semiclassical limit since the s.p. energy does not
enter the VUU equation explicitly, but only its space- and momentum
derivatives. In analogy to the mean-field limit the nucleon phase-space
distribution at $T=0$ is given by a local Thomas-Fermi distribution
(for fixed spin and isospin)
\beq
\label{groud}
f_0({\bf r},{\bf p};t) = \Theta(p_F({\bf r}) - |{\bf p}|)
\eeq
with
\beq
\label{fermi}
p_F({\bf r}) = (\frac{3}{2} \pi \rho({\bf r}))^{1/3}
\eeq
where $\rho({\bf r})$ denotes the nucleon density distribution.

\subsection{Damping of giant quadrupole motion}

Whereas in mean-field theory the frequency of collective modes can
be rather well determined, its damping width appears to be dominated by
higher order $NN$ interactions. As was found in Ref. \cite{r4}
the role of $2p-2h$ matrix elements is very pronounced for giant
quadrupole motion, which we analyse here in more detail.

The time evolution of the quadrupole moment $Q(t)$ and that of $
\dot{Q} (t)$ - which carries a similar information - are
presented in Figs. 2 and 3 for TDHF (upper parts) and TDDM (lower parts) for
an isoscalar quadrupole excitation of 20 MeV of $^{40}Ca$ at an initial
temperature of T=1, 3, and 5 MeV.


\begin{figure}[b]
\centerline{\psfig{figure=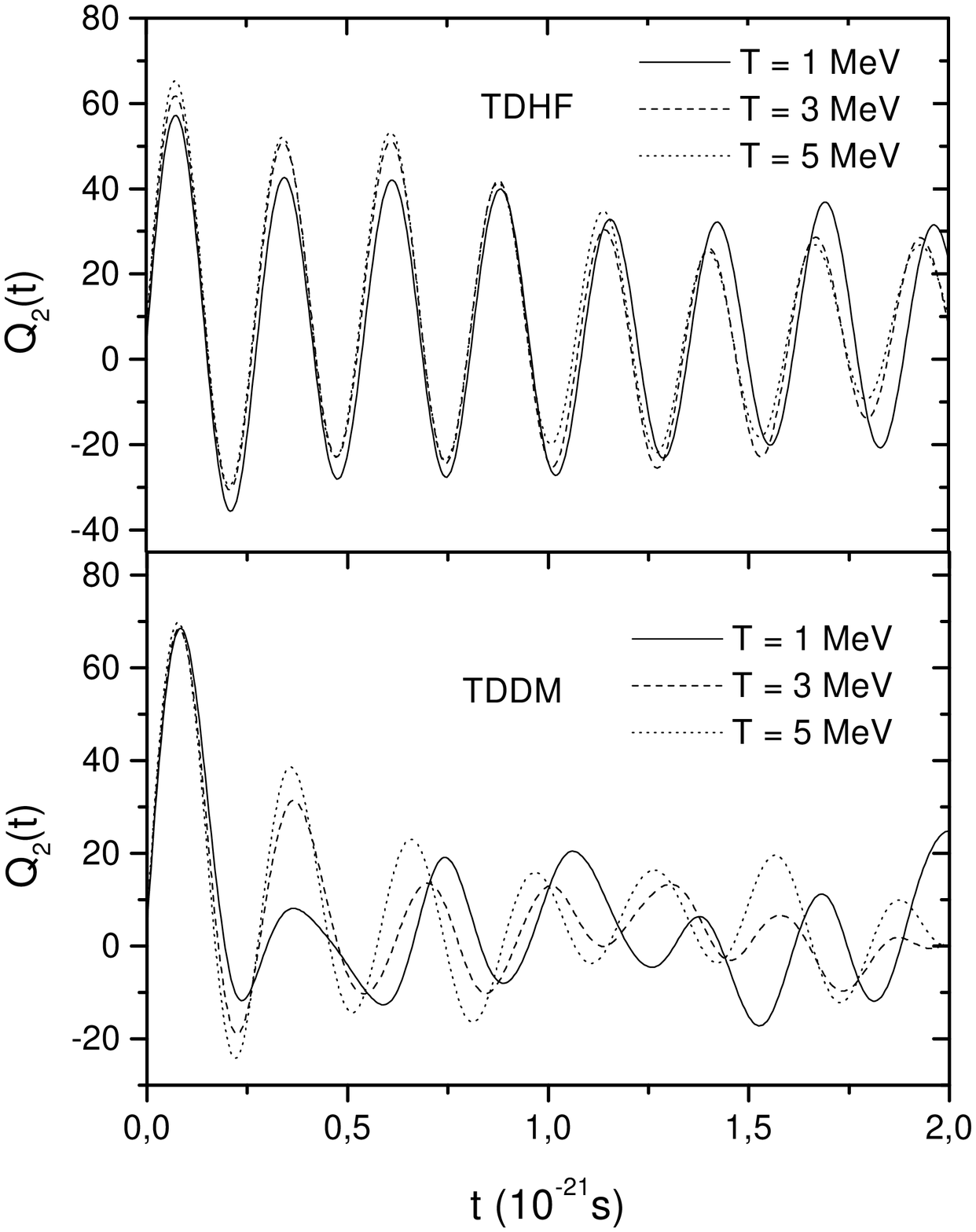,width=10cm}}
\vspace{2.5cm}
{\small Fig.~2: Time dependence of the quadrupole moment $Q_2(t) \equiv
Q(t)$ for an
 isoscalar excitation of $^{40}Ca$ of 20 MeV in the limits TDHF (upper part)
 and TDDM (lower part) at initial temperatures of 1,3, and 5 MeV.}
\label{fig2}
\end{figure}



\begin{figure}[b]
\centerline{\psfig{figure=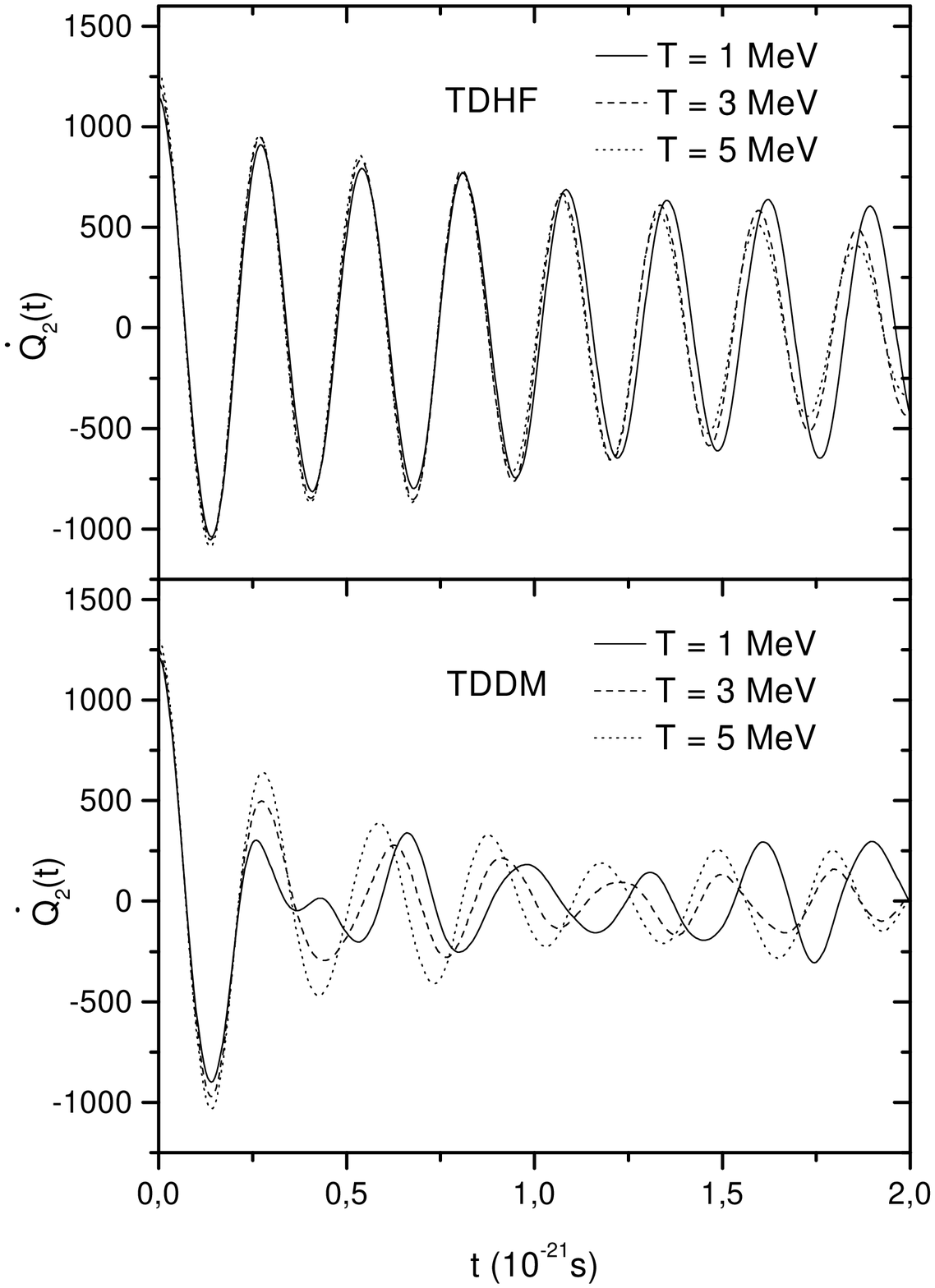,width=10cm}}
\vspace{2cm}
{\small Fig.~3: Time dependence of the quadrupole velocity $\dot{Q}_2 (t) \equiv
\dot{Q} (t)$ for an isoscalar excitation of $^{40}Ca$ of 20 MeV
in the limits TDHF (upper part) and TDDM (lower part)
at initial temperatures of 1, 3, and 5 MeV.}
\label{fig3}
\end{figure}


Whereas there is only small damping
in the case of TDHF,  the collective
motion
is strongly damped in the limit TDDM demonstrating the decisive role of
residual $NN$ interactions.

The questions arises, to what extent the damping of the quadrupole
motion can be attributed to on-shell particle-particle collisions as inherent
in the semiclassical VUU limit. To this aim we show in Fig. 4
the time evolution of the quadrupole moment in momentum space,
\beq
Q_2^P(t) = \frac{4}{(2 \pi)^3} \int d^3r d^3p \
(2p_z^2-p_x^2-p_y^2) f({\bf r},{\bf p};t),
\label{Q2P}\eeq
where $f({\bf r},{\bf p};t)$ results from the VUU equation (\ref{vuu}) in either
the mean-field limit (denoted by Vlasov) or the VUU limit including the on-shell
two-body collisions (VUU). As in case of the quantal approaches there is only a
minor damping in the Vlasov case whereas a rapid decay of $Q_2^P(t)$ is
seen in the VUU limit. Unfortunately, no precise temperature $T$ can be
attributed to the semiclassical calculations due to numerical reasons since
the testparticle method involves numerical fluctuations in the collective
coordinate as well as in the total energy. From the size of the numerical
fluctuations we estimate the 'excitation' temperature to be in the
order of 3-5 MeV.


\begin{figure}[b]
\centerline{\psfig{figure=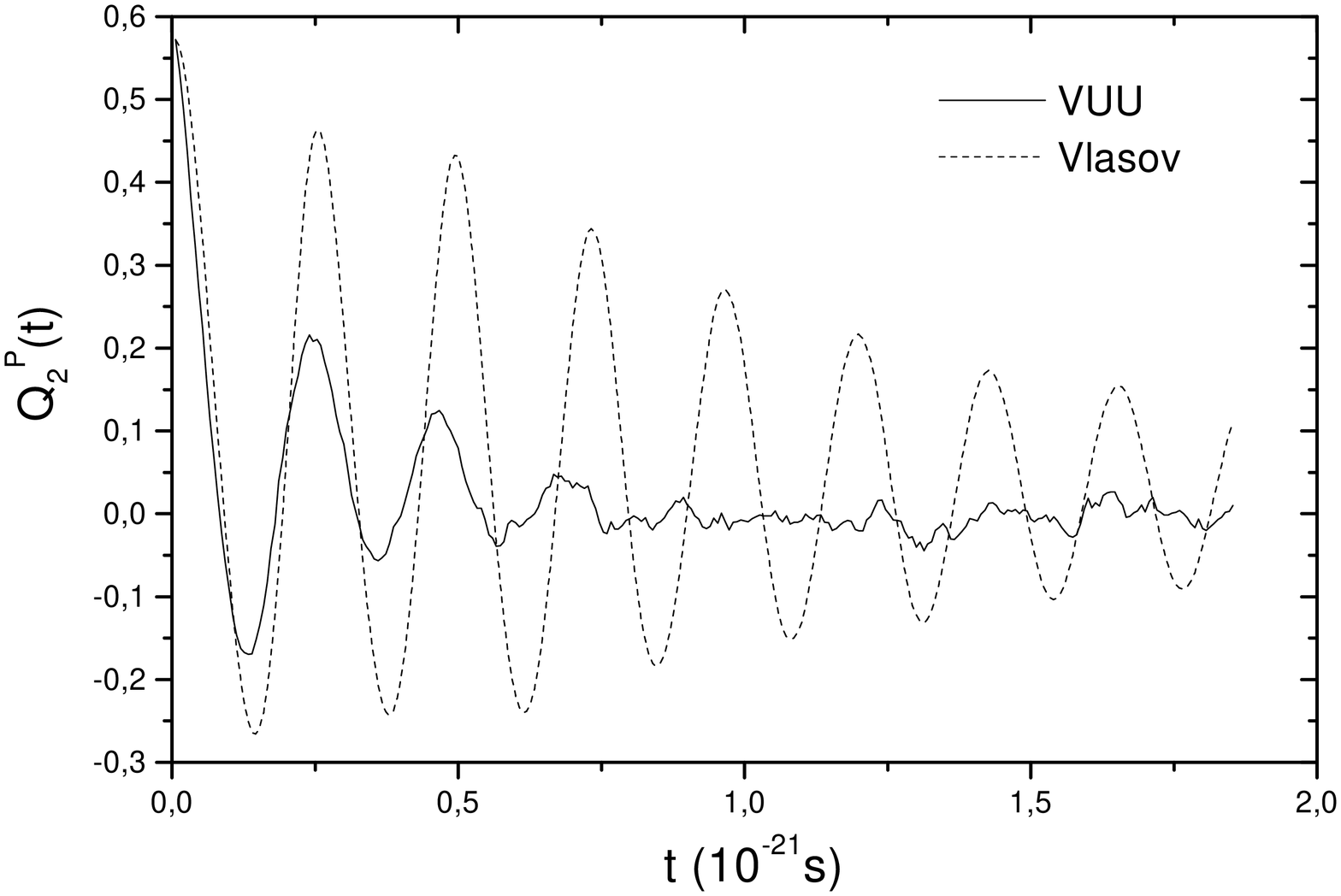,width=10cm,angle=-90}}
\vspace{1.5cm}
{\small Fig.~4: Time evolution of the quadrupole moment in momentum space (42)
in the limits VUU (solid line) and Vlasov (dashed line)}
\label{fig4}
\end{figure}


Since the VUU approach includes only on-shell particle-particle collisions and
leaves out particle-hole interactions, which are important for the
damping of shape vibrations \cite{gfb83}, it yields a moderate damping in
between the limits TDHF and TDDM.
From the height and position of the
maxima (and minima) we can extract the average collective frequency
 $\Omega$ in all cases as well as an average friction constant
$\Gamma_{Q}$ using $Q(t)\sim \exp(-\Gamma_{Q}t/2\hbar)$. The
latter quantity is shown in Fig. 5 for initial nuclear temperatures T
of 0, 1, 2, 3, 4, 5, and 6 MeV for TDHF and TDDM where the error bars
indicate the uncertainty of extraction from the time-dependent signal
$Q(t)$ or $\dot{Q}(t)$.


\begin{figure}[b]
\centerline{\psfig{figure=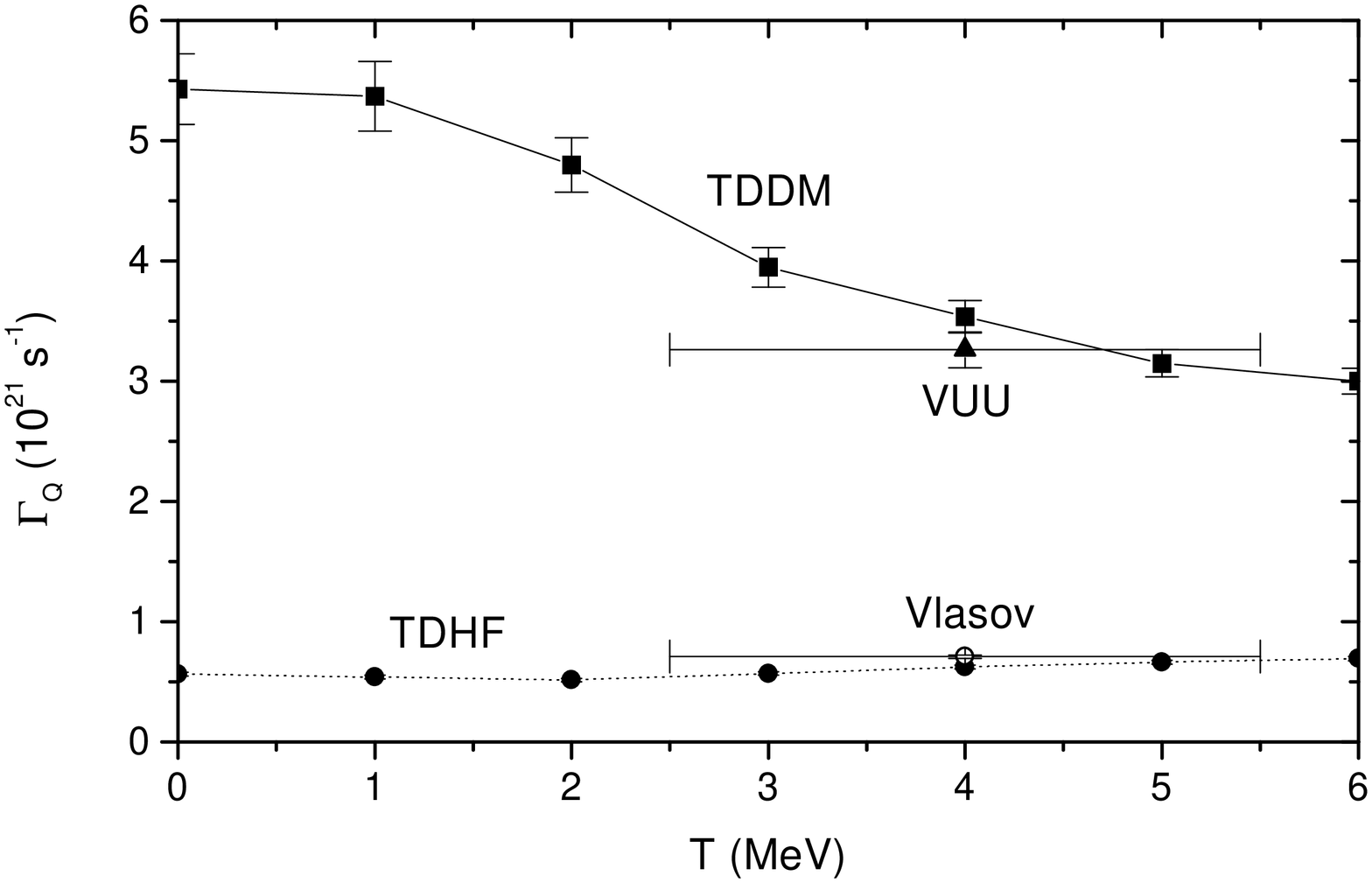,width=10cm,angle=-90}}
\vspace{1.5cm}
{\small Fig.~5: Extrapolated values  for the friction coefficient $\Gamma_Q$ as a
function of the initial temperature $T$ in the limits TDDM, TDHF, VUU, and Vlasov.}
\label{fig5}
\end{figure}


The actual values are found to be approximately
independent on temperature for
TDHF, whereas the damping drops with $T$
for TDDM.  Fig. 5, furthermore, shows that the damping achieved in TDHF
compares well with the semiclassical Vlasov limit for $T\simeq 3-5$~MeV
and the additional damping introduced by collisions in VUU only
provides about half of the friction obtained from TDDM at low temperature
(T $\approx$ 0).  This result
indicates the essential role played by the  particle-hole correlations
(long-range correlations) in the damping of shape and surface
vibrations.  As discussed in \cite{r4,s2} a sizeable damping arizes
 from the particle-hole (p-h) interaction matrix $H_{\alpha \beta
\alpha' \beta'}$ (\ref{2.11}) that is not considered in the
limit with nonperturbative
p-p and h-h scattering (e.g. VUU like models or in Ref. \cite{28b}).

The actual magnitude of $\Gamma_Q \approx 5-6 \cdot 10^{21} s^{-1}$ at
$T \approx $0 from TDDM
compares well with the friction constant extracted experimentally
from the width of the GQR for $^{40}Ca$
and indicates the importance of p-h residual matrix elements.

The dropping of the width $\Gamma_Q$ with temperature for TDDM in case of
isoscalar quadrupole motion can be attributed to the fact that the long-range
correlations decrease in magnitude with temperature. In order to demonstrate
this dependence we show in Fig. 6 the dependence of the correlation energy
on temperature $T$ which in magnitude decreases by a factor of 2 from $T$ = 0
to $T$ = 6 MeV. Thus the 'long-range correlations melt with temperature' and
the width of the giant quadrupole resonances should decrease with
$T$. This is opposite to the role of on-shell 2-body collisions which are
blocked in the VUU limit at T=0 and increase with phase space as $\sim T^2$.
Off-shell collisions allow for a faster
increase of phase-space with temperature
as demonstrated in Ref. \cite{28b}, however, the functional dependence of
the collisional width depends crucially on the multipolarity of the mode.
Thus dominantly the particle-hole correlations induced by (\ref{2.11})
generate sizeable density fluctuations at low temperature
T which lead to a coupling of
different collective modes and a redistribution of their strength.


\begin{figure}[b]
\centerline{\psfig{figure=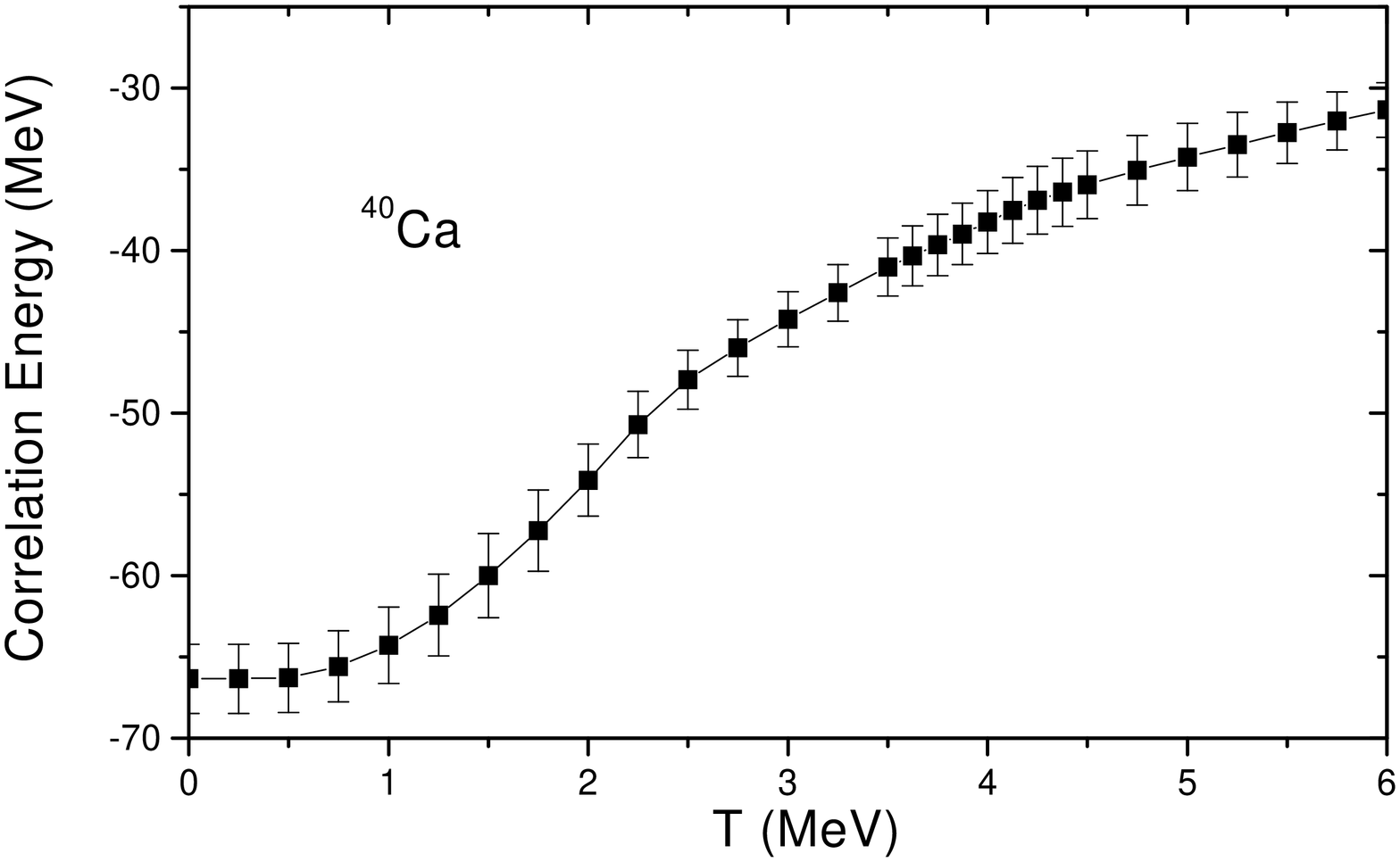,width=10cm,angle=-90}}
\vspace{2cm}
{\small Fig.~6: The correlation energy for $^{40}Ca$ in the limit TDDM as a
function of temperature $T$.}
\label{fig6}
\end{figure}

\subsection{Dispersion of giant quadrupole motion }
In the quantum theories the dispersion of one-body operators can be evaluated
explicitly and the effect of 2-body correlations on these 2-body
operators can be analyzed in detail. We note that
a collective mass parameter can be extracted from the calculations via
\beq
\label{3.1}
E_{coll} \approx {M_Q \over {2}} \langle \dot{Q}\rangle^2_{max} \approx {M_Q \over {2}}
\Omega^2 \langle Q\rangle^2_{max}
\eeq
and leads to $M_Q \approx 2.0 {MeV / ({fm^2 c^2})}$ which is practically
equivalent
with the value of the direct calculation from Eq. (\ref{2.15}).
The calculated
dispersions $\sigma_Q^2(t)$ and $\sigma_{\dot{Q}}^2(t)$ according to
(\ref{2.16}) are found
to oscillate in phase with the quadrupole
moment $\langle Q\rangle (t)$ or $\langle \dot{Q}\rangle (t)$ (cf. Fig.7).
When using the same initial conditions $\sigma_{\dot{Q}}^2(t=0)$ and
$\sigma_Q^2(t=0)$ as well as the collective frequency $\Omega$ and $\gamma_Q=
\Gamma_Q$ from the TDDM calculations, the quantum harmonic oscillator
(QHO, Eqs. (33)-(35)) gives comparable solutions for $\sigma_Q^2(t)$ and $\sigma_{\dot{Q}}^2(t)$
which, however, are damped out faster in time (dotted lines in Fig. 7). In
case of TDDM or TDHF the solutions still oscillate around some average value.


\begin{figure}[b]
\centerline{\psfig{figure=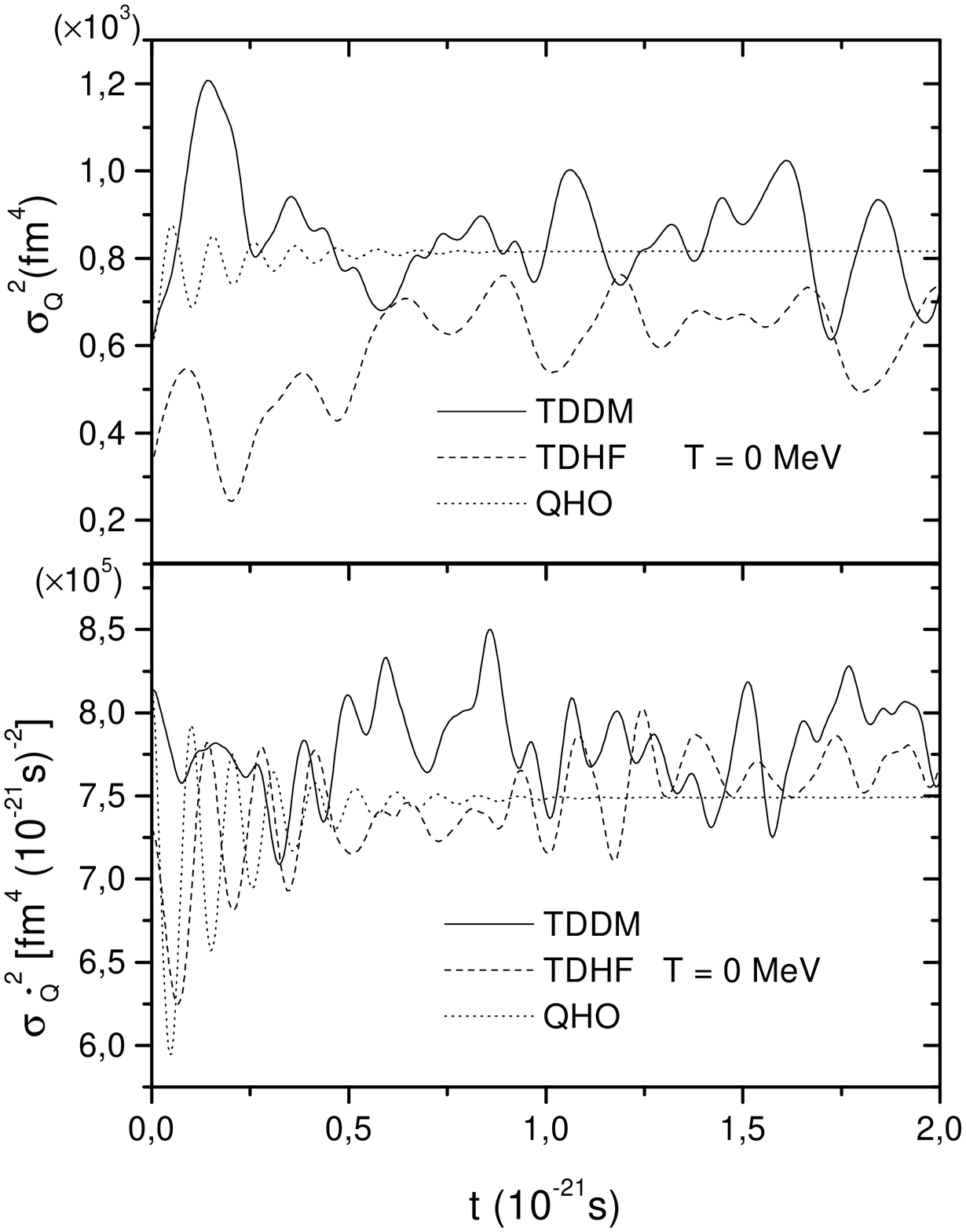,width=10cm,angle=0}}
\vspace{2.5cm}
{\small Fig.~7: The fluctuations
$\sigma_Q^2(t)$ (upper part) and $\sigma_{\dot{Q}}^2(t)$ (lower part)
for $^{40}Ca$ excited collectively by 20 MeV in the limit TDDM (full lines)
and TDHF (dashed line) in comparison to the results from the damped
quantum harmonic oscillator (QHO; dotted lines).}
\label{fig7}
\end{figure}

 The asymptotic values of $\sigma_Q^2$ and
$\sigma_{\dot{Q}}^2$ (for $t \rightarrow \infty$)
are obtained by averaging over the final time interval
(typically for $t \geq 7.5 \cdot 10^{-22} sec$) and shown in Fig. 8
 as  functions
of the initial temperature $T$. Whereas $\sigma_Q^2$ and $\sigma_{\dot{Q}}^2$
only slightly varies with temperature for the QHO, the variation with $T$ is
more pronounced for TDDM (solid lines) and especially TDHF (dashed lines),
which reflect the intrinsic dynamics. Thus model studies based on the
QHO equations of motion should be taken with care especially at high
temperature \cite{35c}.


\begin{figure}[b]
\centerline{\psfig{figure=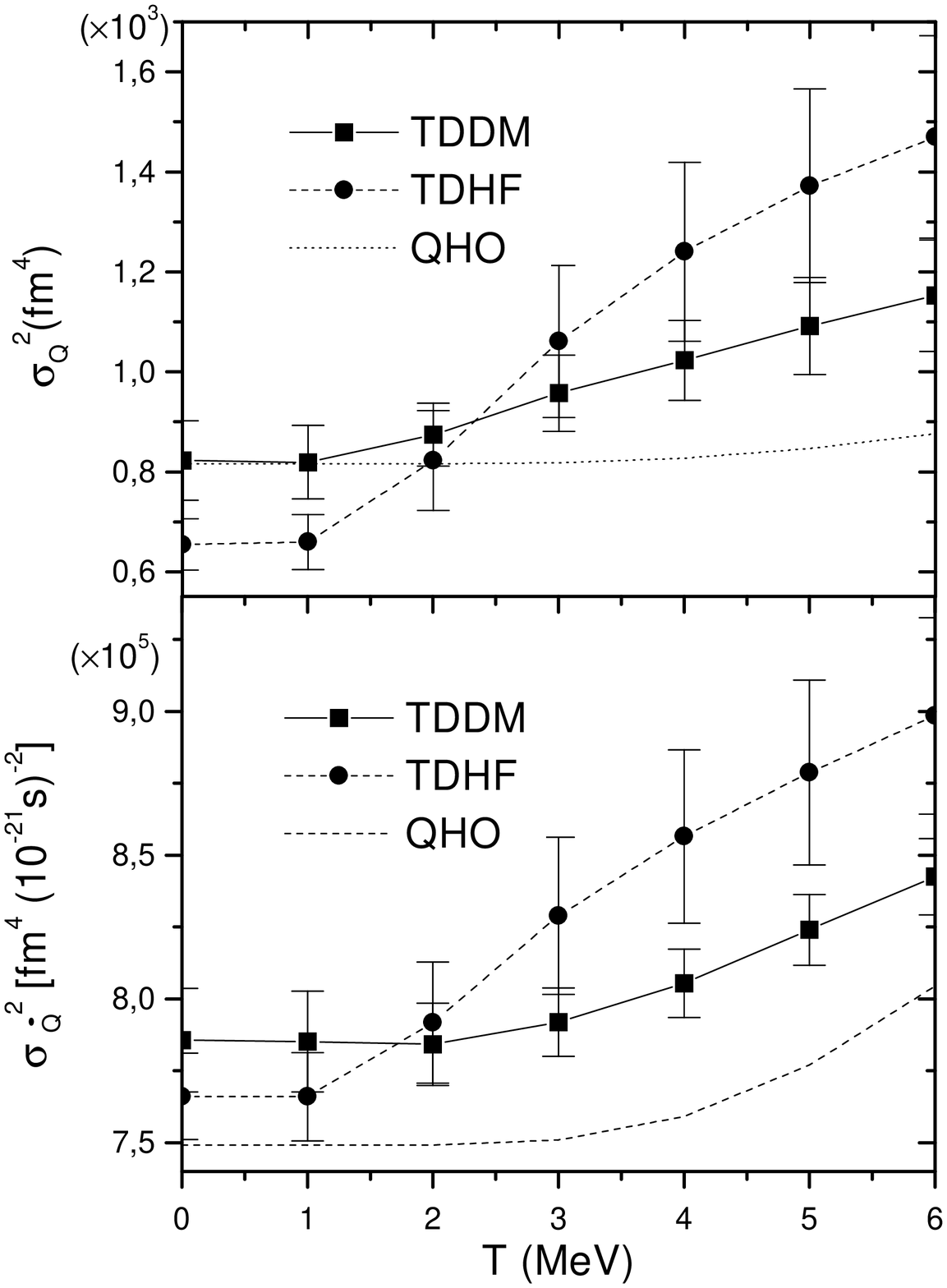,width=10cm,angle=0}}
\vspace{2.5cm}
{\small Fig.~8: Asymptotic fluctuations in the quadrupole degree of freedom
$\sigma_Q^2(\infty)$ (upper part) and in the quadrupole velocity
 $\sigma_{\dot{Q}}^2(\infty) $ (lower part)
in the limits TDDM, TDHF and the quantum harmonic oscillator QHO.}
\label{fig8}
\end{figure}


The product $M_Q^2 \sigma_{\dot{Q}}^2
\sigma_Q^2$ according to (\ref{2.15b}) must obey the uncertainty relation
in all quantum mechanical approaches. This is demonstrated in Fig. 9
for TDDM (solid line), TDHF (dashed line) and the QHO (dotted line) as
a function of temperature $T$. Note, that $\hbar^2 \approx 0.4 \ (MeV \
10^{-21}s)^2$.

\begin{figure}[b]
\centerline{\psfig{figure=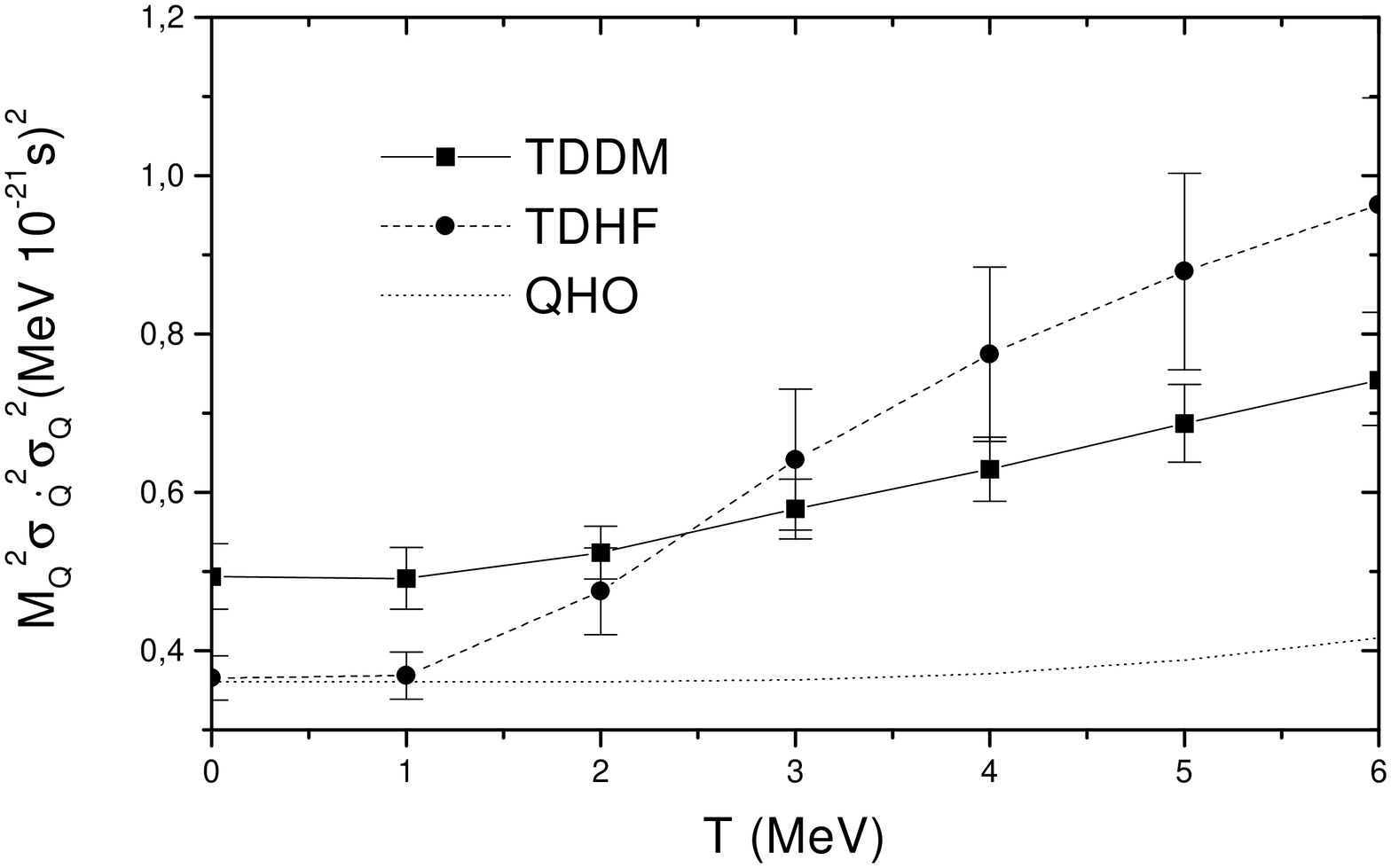,width=10cm,angle=-90}}
\vspace{2.5cm}
{\small Fig.~9: Test of the uncertainty relation for TDDM (solid line), TDHF (dashed line) and the QHO (dotted line) as a function of $T$.}
\label{fig9}
\end{figure}



\begin{figure}[b]
\centerline{\psfig{figure=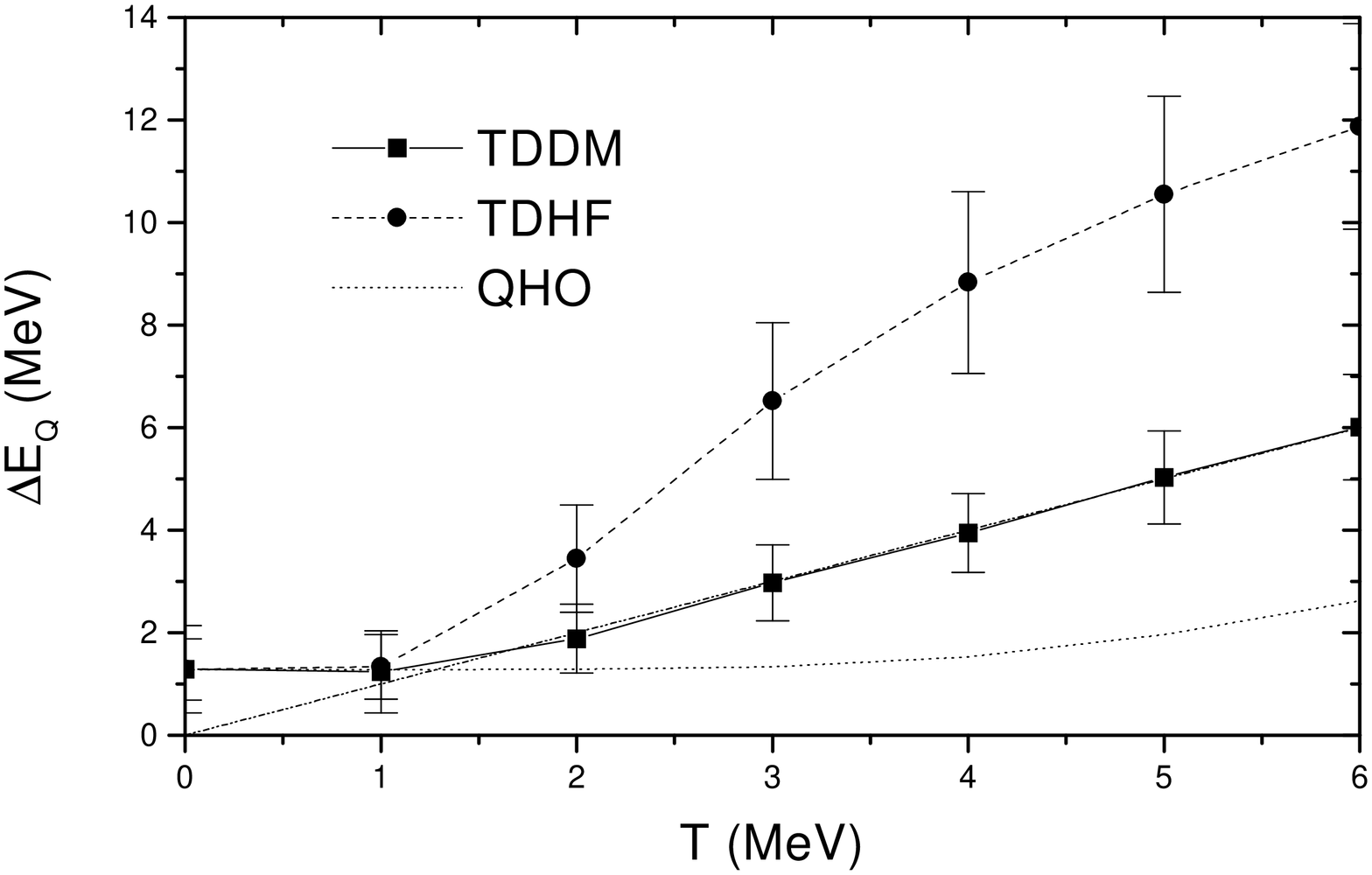,width=10cm,angle=-90}}
\vspace{2.5cm}
{\small Fig.~10: Test of the equipartition relation for $\Delta E_Q(T)$
(45) (straight dash-dotted line)
for the collective
fluctuations as a function of temperature
in the limits TDDM, TDHF and the quantum harmonic oscillator QHO.}
\label{fig10}
\end{figure}


Since $\frac{1}{2}M_Q\sigma_{\dot{Q}}^2 $
 together with the potential energy $\frac{1}{2}M_Q\Omega^2\sigma_{Q}^2$
should reflect the equipartition theorem in case of damped harmonic motion,
\cite{20d}
\beq
\label{3.2}
E_Q(T) = {1 \over {2}} M_Q \sigma_{\dot{Q}}^2+
{1 \over {2}}M_Q \Omega\sigma_{Q}^2 = E_Q(0) + T  ,
\eeq
and thus be a linear function of the
temperature $T$ according to Eq. (\ref{3.2}),
we have plotted
\beq
\label{Ediff}
\Delta E_Q(T) = E_Q(T)-E_Q(0)
\eeq
in Fig. 10 for all three cases.
Here only the full two-body theory TDDM follows equipartition for
$T \geq$ 1 MeV, i.e. $\Delta E_Q(T) = T$,
whereas TDHF as well as the QHO significantly overestimate or underestimate the
dash-dotted line, respectively. The off-set of $\approx$ 1 MeV for
TDDM at $T=0$ so far is not understood.

\section{Specific heat, entropy and free energy}

Apart from the crucial role of long-range correlations in the damping
of the collective quadrupole motion demonstrated in Section 3, the
two-body correlations are expected to modify also the thermodynamic
properties relative to a mean-field approach. To investigate this question
we calculate the total energy of our test-nucleus $^{40}Ca$ (without any
collective excitation) as a function of the initial temperature from
$T$= 0 to $T$ = 6 MeV. We note that due to two-body correlations the
binding energy is lowered by $\approx$ 20 MeV (i.e. -382 MeV) for TDDM as
compared to TDHF (- 362 MeV).

In Fig. 11 we show the total energy for TDDM (full squares; shifted by 20 MeV)
and TDHF (full circles) as a function of temperature (upper part) as well as
the specific heat (\ref{2.20a}) (lower part). The total energy increases
slightly faster with $T$ for TDDM as for TDHF due to a decrease in the
correlation energy with $T$ (cf. Fig. 6). Whereas the increase of the
energy is monotonous for TDHF, there is a more rapid variation of the
total energy for TDDM at $T \approx$ 4 MeV. This is seen more clearly in the
specific heat (lower part), where $C_V$ shows a sudden bump around $T \approx$ 4 MeV,
whereas the specific heat is flat as a function of $T$ for TDHF.


\begin{figure}[b]
\centerline{\psfig{figure=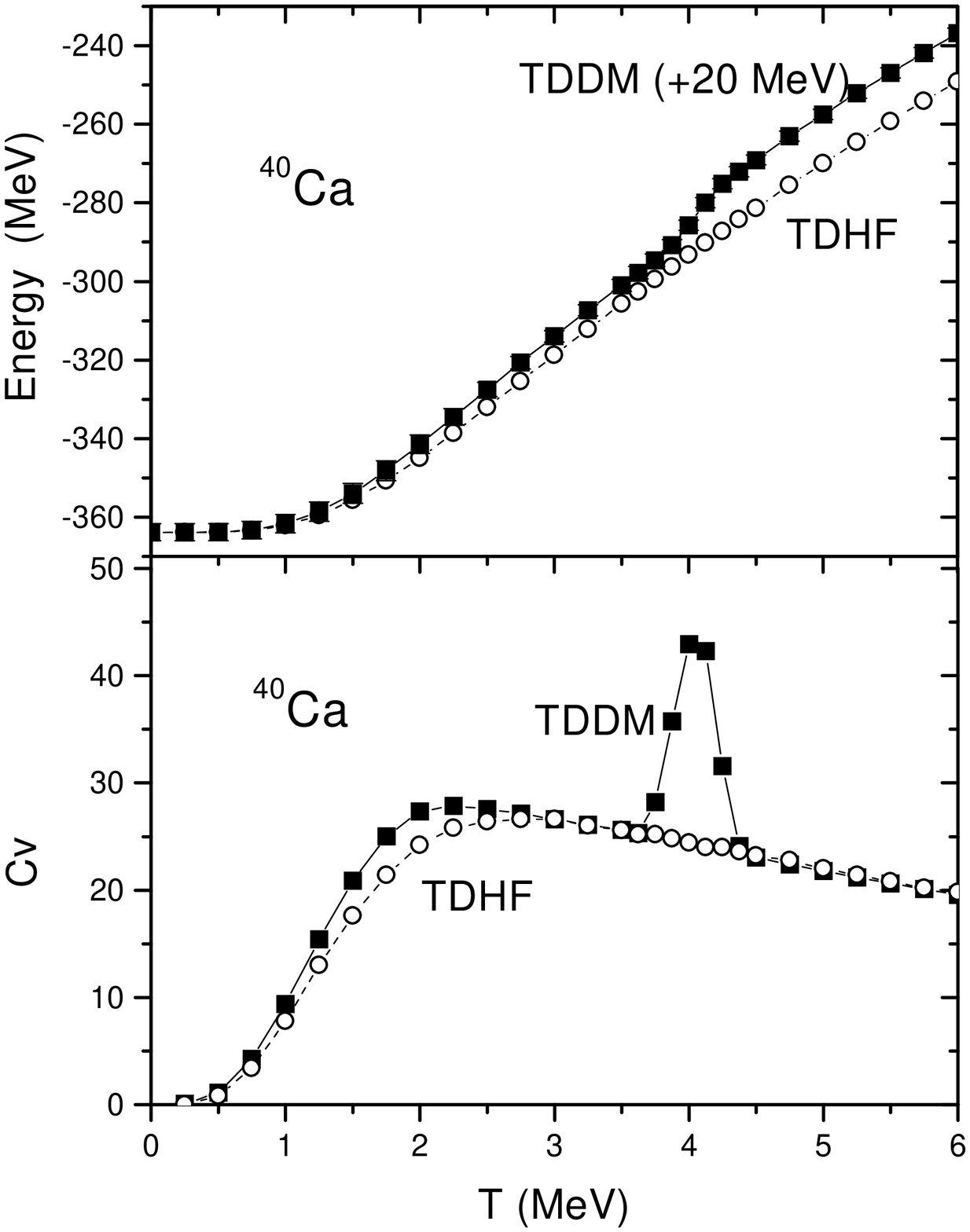,width=10cm,angle=0}}
\vspace{2.5cm}
{\small Fig.~11: The total energy (upper part) and the specific heat (27)
(lower part) of the system $^{40}Ca$ as a function of temperature
 in the limits of TDHF (open circles) and TDDM (full squares).}
\label{fig11}
\end{figure}


Since conventionally maxima in the specific heat are attributed to
phase transitions, where the fluctuations in energy become very large,
we have evaluated the quantity $<H^2> - <H>^2$ according to (\ref{2.21}). As
seen from Fig. 12 the fluctuations in the total energy also show a bump
at $T \approx$ 4 MeV for TDDM (solid triangles) whereas they stay rather
flat for TDHF (open circles) in this temperature region.


\begin{figure}[b]
\centerline{\psfig{figure=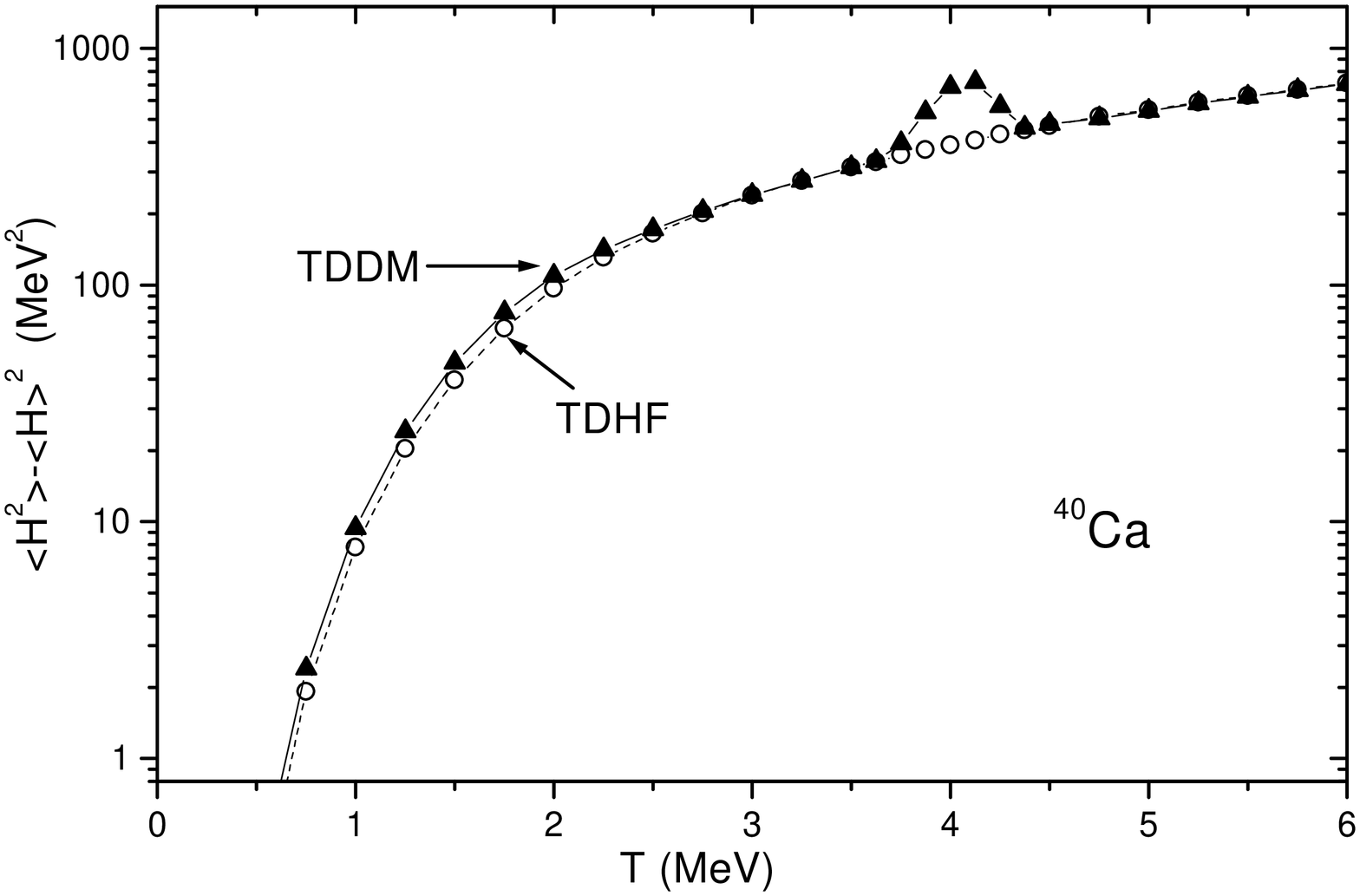,width=10cm,angle=-90}}
\vspace{2.5cm}
{\small Fig.~12: The fluctuation in energy (30)
 of the system $^{40}Ca$ as a function of temperature
 in the limits of TDHF (open circles) and TDDM (full triangles).}
\label{fig12}
\end{figure}


The origin of this bump is not due to a rapid change of the correlation
energy (cf. Fig. 6) but reflects the melting of single-particle effects in
case of TDDM. This is demonstrated in Fig. 13 where we show the single-particle density
$\rho(x=y=0;z)$ as a function of the coordinate $z$ for temperatures
from $T$=0 to $T$ = 6 MeV (upper part) as well as the single-particle
potential (lower part). The pronounced maxima (for $z \approx$ 0) in case
of low temperature $T \leq $ 3 MeV no longer appear for $T \geq$ 4 MeV contrary to
the case of TDHF. We attribute this behaviour to the melting of shell structure
by the two-body correlations in case of TDDM which is accompanied by a larger
specific heat as well as larger fluctuations in the total energy (cf.
Figs. 11 and 12).


\begin{figure}[b]
\centerline{\psfig{figure=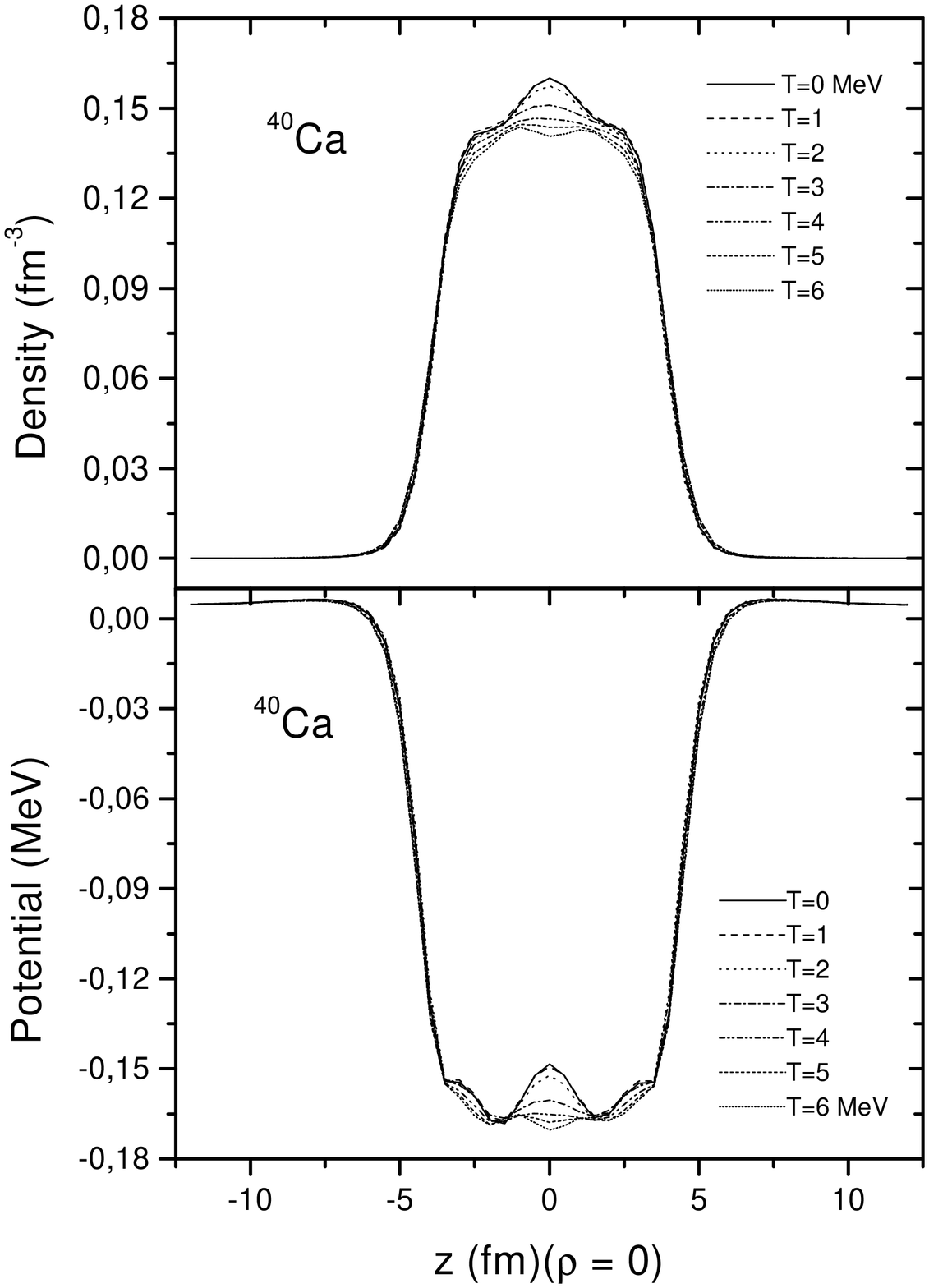,width=10cm,angle=0}}
\vspace{2.5cm}
{Fig.~13: The nuclear density (upper part) and the
single-particle  potential (lower part) for $^{40}Ca$
   as a function of the coordinate $z$ for temperatures $T$= 1,2,3,4,5,6 MeV
 in the limit   TDDM.}
\label{fig13}
\end{figure}

We finally note that the role of two-body correlations also shows up in
the entropy $S$ (\ref{2.20b}) and the free energy $F$ (\ref{2.20c}) of the
nucleus. The latter two quantities are displayed in Fig. 14 as a function of
temperature for TDDM and TDHF. Whereas the two-body correlations lead to a
decrease of the free energy $F$ relative to the mean-field result,
they enhance the entropy of the system
due to a stronger dynamical mixing of single-particle configurations
(Slater-determinants). However, the relative changes turn out to be
quite moderate except for a shift of the free energy by $\approx$ 20 MeV
for TDDM due to the correlation energy.

\begin{figure}[b]
\centerline{\psfig{figure=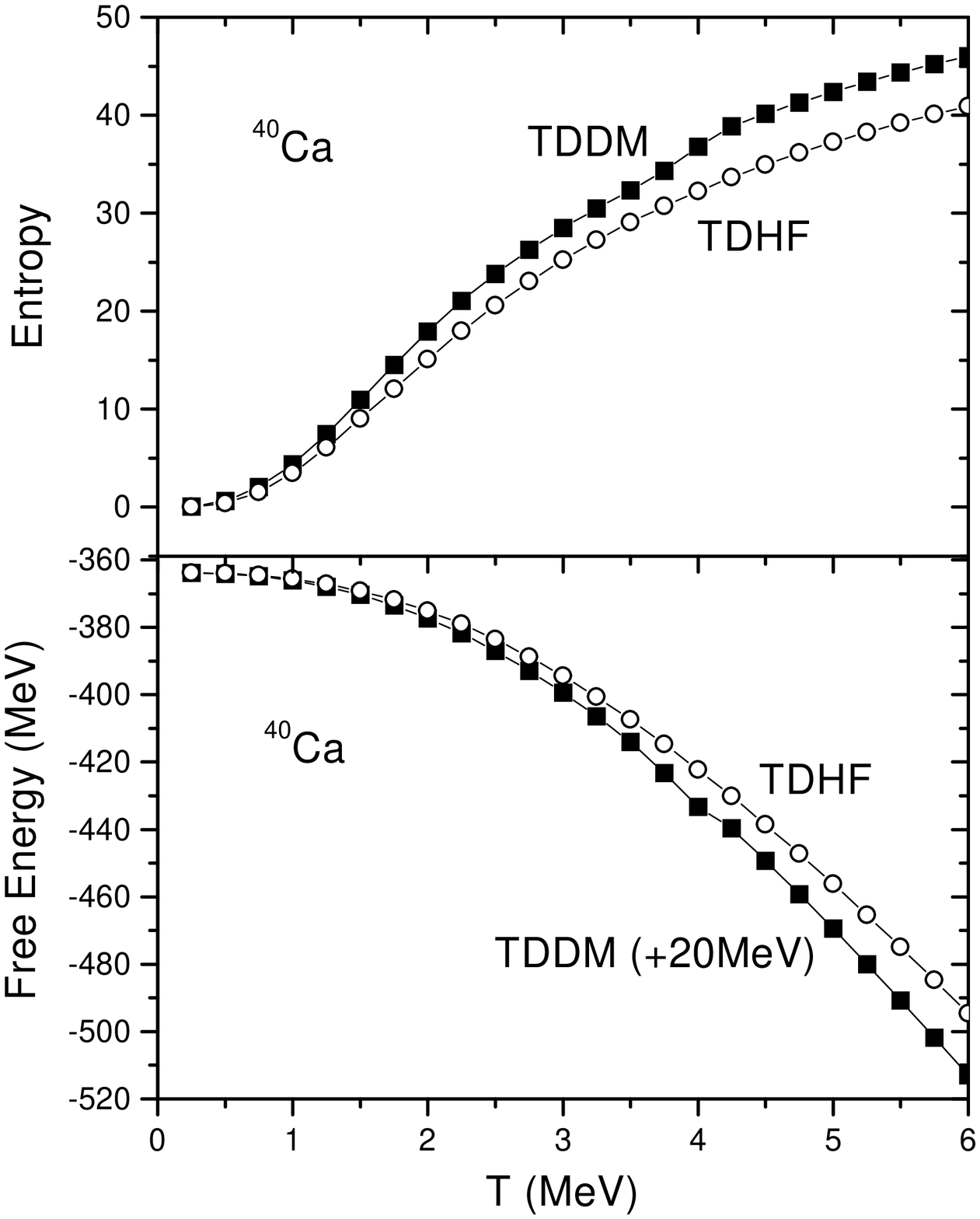,width=10cm,angle=0}}
\vspace{3cm}
{\small Fig.~14: The entropy $S$ (28) (upper part) and free energy $F$ (29)
(lower part) for $^{40}Ca$
   as a function of temperature for   TDDM and TDHF.}
\label{fig14}
\end{figure}

\section{Summary and discussion}

In this work we have presented microscopic calculations for the
nucleus $^{40}Ca$ within a nonperturbative two-body correlation
approach which
superseedes TDHF as well as the 'traditional' VUU model \cite{r1,r2}
by including  explicit p-h interactions
as well as off-shell collision processes to all orders. Note that more recent
extensions of mean-field theory \cite{28b,35b} include off-shell collisions,
however, are still restricted to the Born approximation.

By studying the decay of
collective quadrupole excitations we have extracted
transport coefficients for friction
and diffusion and analyzed the fluctuation properties of the theory which
fulfills the
equipartion theorem for the fluctuations in the collective velocity
(for $T \geq$ 1 MeV) contrary to the mean-field limit TDHF.
This study has also shown the essential role played
by the particle-hole (long-range) correlations in the damping
of shape and surface vibrations in nuclei at lower temperature whereas
for $T \geq$ 3-4 MeV the particle-particle collisions take over and the
system might be conveniently described within the semiclassical VUU limit.
The latter phenomena is due to a decrease of the correlation energy with
temperature and an increase with phase-space of on-shell two-body collisions
with $T$.

The temperature dependence of the entropy $S$ and the free energy $F$
is found not to be modified substantially due to
two-body correlations relative to the
mean-field limit. However, a small bump in the specific heat $C_V$ and
in the fluctuations in energy shows up for $T \approx$ 4 MeV, which does not
appear in mean-field theory. We attribute this effect to a melting of shell structure
due to two-body correlations.

In view of the detailed microscopic analysis we expect that nucleus-nucleus
collisions above 20 A MeV should be adequately described in the semiclassical
VUU limit as far as stopping power and single-particle observables are
concerned. Multifragmentation phenomena, however, might require the dynamics
of long-range correlations \cite{r30,r17} especially at low nuclear density.
Mean-field dynamics here provide a glance at the complexity of the shape
evolution \cite{Jung,38c} but should not be adequate to describe nuclear
condensation phenomena or the growth of instabilities according to the
analysis in this work. First steps in exploiting Eqs.
(1) - (4) for the description of nucleus-nucleus collisions have been made
in Ref. \cite{wang}.

\end{document}